\documentclass[journal]{IEEEtran}
\usepackage{cite}
\ifCLASSINFOpdf
\usepackage[pdftex]{graphicx}
\DeclareGraphicsExtensions{.pdf,.jpeg,.png}
\else
\usepackage[dvips]{graphicx}
\DeclareGraphicsExtensions{.eps}
\fi
\usepackage{epsfig,epsf,epstopdf,graphicx}
\usepackage{caption}
\usepackage[cmex10]{amsmath}
\usepackage{amssymb}
\usepackage{amsthm }
\usepackage{nicefrac}
\usepackage{hyperref}
\usepackage{xcolor}
\usepackage{bm}
\hypersetup{colorlinks=true}
\usepackage{tabularx}

\theoremstyle{theorem}

\theoremstyle{definition}

\theoremstyle{plain}

\theoremstyle{plain}

\newcommand{\?}[1]{\kern-.#1em }

\newcommand{\xb}{{\bm{x}}}

\newcommand{\Bb}{{\bm{B}}}

\newcommand{\sbb}{{\bm{s}}}

\newcommand{\ub}{{\bm{u}}}
\newcommand{\Ub}{{\bm{U}}}

\newcommand{\eb}{{\bm{e}}}

\newcommand{\bb}{{\bm{b}}}

\newcommand{\zb}{{\bm{z}}}

\newcommand{\Ib}{{\bm{I}}}

\newcommand{\vb}{{\bm{v}}}

\newcommand{\Lb}{\mathbf{L}}

\newcommand{\Mb}{\mathbf{M}}
\usepackage{algorithm}
\usepackage{multirow}
\usepackage{algcompatible}
\usepackage[bottom]{footmisc}
\PassOptionsToPackage{dvipsnames}{xcolor}
\usepackage{tikz}
\usetikzlibrary{calc}

\ifCLASSOPTIONcompsoc
  \usepackage[caption=false,font=normalsize,labelfont=sf,textfont=sf]{subfig}
\else
  \usepackage[caption=false,font=footnotesize]{subfig}
\fi
\usepackage{fixltx2e}
\usepackage{afterpage}
\usepackage{float}
\usepackage{stfloats}

\makeatletter

\begin{document}

\title{Secure Blind Graph Signal Recovery and Adversary Detection Using Smoothness Maximization }

\author{Mahdi~Shamsi,
Hadi Zayyani,
Hasan Abu Hilal,
and Mohammad Salman

\thanks{M.~Shamsi and H.~Zayyani are with the Department
of Electrical and Computer Engineering, Qom University of Technology (QUT), Qom, Iran (e-mails: mahdi.shamsi@alum.sharif.edu, zayyani@qut.ac.ir).}
%\thanks{H.~Zayyani is with the Department
%of Electrical and Computer Engineering, Qom University of Technology (QUT), Qom, Iran (email: zayyani@qut.ac.ir).}% <-this %
\thanks{H.~Abu Hilal, is with the Electrical Engineering Department, Higher Colleges of Technology, Abu Dhabi, UAE (email: hasan.abuhilal@hct.ac.ae).}
\thanks{M.~Salman is with College of Engineering and Technology, American University of the Middle East, Egaila 54200, Kuwait (e-mail: mohammad.salman@aum.edu.kw).}}% <-this %


\maketitle


\begin{abstract}
In this letter, we propose a secure blind Graph Signal Recovery (GSR) algorithm that can detect adversary nodes. Some unknown adversaries are assumed to be injecting false data at their respective nodes in the graph. The number and location of adversaries are not known in advance and the goal is to recover the graph signal in the presence of measurement noise and False Data Injection (FDI) caused by the adversaries. Consequently, the proposed algorithm would be a perfect candidate to solve this challenging problem. Moreover, due to the presence of malicious nodes, the proposed method serves as a secure GSR algorithm. For adversary detection, a statistical measure based on differential smoothness is used. Specifically, the difference between the current observed smoothness and the average smoothness excluding the corresponding node. This genuine statistical approach leads to an effective and low-complexity adversary detector. In addition, following malicious node detection, the GSR is performed using a variant of smoothness maximization, which is solved efficiently as a fractional optimization problem using a Dinkelbach's algorithm. Analysis of the detector, which determines the optimum threshold of the detector is also presented. Simulation results show a significant improvement of the proposed method in signal recovery compared to the median GSR algorithm and other competing methods.

\end{abstract}

\begin{IEEEkeywords}
Graph signal, recovery, adversary, secure, smoothness.
\end{IEEEkeywords}

% \fi
\IEEEpeerreviewmaketitle

\section{Introduction}
\label{sec:Intro}
\IEEEPARstart{G}{raph} Signal Processing (GSP) is an emerging paradigm in signal processing where signals are defined over irregular domains, such as graphs. GSP has found wide-ranging applications in areas including biological networks, transport systems, image processing, and biomedical signal analysis \cite{Shum13}-\cite{Ortega22}. The literature on GSP addresses several fundamental problems, such as graph signal recovery (GSR), graph signal sampling, graph learning, and graph signal representation \cite{Shum13}-\cite{Ortega22}.

In GSR, the objective is to reconstruct the original graph signal from only a subset of its sampled values, assuming that the sampling matrix is known. Existing GSR methods can be broadly categorized into non-adaptive \cite{Chen15}-\cite{Tork21} and adaptive approaches \cite{Lore16}-\cite{Tork25}. Adaptive approaches often rely on distributed estimation, where a set of local estimators collaborate to achieve a common inference task. Within this framework, various strategies have been proposed, including interference cancellation \cite{flexible}, joint distributed estimation with mask learning \cite{jointMask, maskRetieve, sparseMask}, recovery under partial observations with impulsive noise \cite{impNoise}, and clustered multi-task diffusion, where nodes in each cluster pursue related but not identical goals compared to neighboring clusters \cite{MTD}.

This paper, however, focuses primarily on non-adaptive methods. In this category, a range of algorithms have been developed, such as variational minimization \cite{Chen15}, distributed tracking \cite{Wang15}, local-set-based algorithms \cite{Wang15may}-\cite{Wang16}, smoothness-based methods \cite{Qiu17}-\cite{Qiu18}, primal-dual optimization \cite{Berg17}, kernel-based techniques \cite{Rome17}, autoregressive model-based algorithms \cite{Ioan19}, posterior recovery of diffused sparse signals \cite{Rey19}, truncated Neumann series methods \cite{Wang18}, iterative recovery approaches \cite{Brug20}, and a Variational Bayes (VB) framework \cite{Tork21}.

\noindent In most of the existing algorithms, measurement noise is considered as Gaussian noise and the sampling nodes are known in advance. In some other works, impulsive noise is considered in the GSR problem \cite{Tork22}-\cite{Yama25}. In addition, security has become an important issue in nowadays systems \cite{Yang20}. To best of our knowledge, this paper is the first to introduce the secure graph signal recovery. We assume that some unknown nodes in the graph inject  arbitrary malicious signal values of the graph signal. Hence, the objective is to robustly recover the graph signal in the presence of such adversary nodes. Since these adversary nodes are unknown, the GSR problem is indeed a blind GSR problem, and because the presented algorithm is robust to adversaries, this makes it a secure GSR algorithm. Moreover, in this paper, a low complexity adversary detector is proposed based on the smoothness assumption of the graph signal, which is prevalent in GSP \cite{Qiu17}-\cite{Qiu18}, \cite{Tork25}. Once adversarial nodes are detected, a smoothness-based optimization algorithm is applied for GSR. This leads to a fractional optimization problem, which is efficiently solved using Dinkelbach’s algorithm \cite{dinkelbach1967nonlinear}.
Moreover, a theoretical analysis of the proposed detector is provided and an optimum threshold for the detector is obtained. Simulation results show an improvement over the median filtering algorithm and other graph-based denoising algorithms \cite{Tay21}-\cite{Tay21SP}.

\section{System Model and Problem Formulation}
\label{sec: prob}
\noindent Consider a graph $\mathcal{G} = (\mathcal{V},\mathcal{E})$ which is composed of $N$
nodes $\mathcal{V}=\{1,2,\dots,N\}$ and a set of weighted edges $\mathcal{W}=\{w_{ij}\}_{i,j} \subset \mathcal{V}\times \mathcal{V}$. The weight matrix $\mathbf{W}$ contains the weights $w_{ij}$, while the degree of a node $i$  is defined as $d_i = \Sigma_{j=1}^{N} w_{ij}$. Collecting these degrees in a diagonal matrix $\mathbf{D}$, the  Laplacian of the graph $\mathbf{L}$ is given by $ \mathbf{L} = \mathbf{D} -\mathbf{W}$. For undirected graphs, the Laplacian $\mathbf{L}$  is symmetric and positive semi-definite, allowing eigendecomposition $\mathbf{L} = \mathbf{U}\mathbf{\Lambda}\mathbf{U}^\top$, where $\mathbf{U}$ contains the eigenvectors and $\mathbf{\Lambda}$ is the diagonal matrix of eigenvalues. This decomposition is central to the spectral analysis of graphs \cite{Ortega22}. Analogously to classical signal processing, the graph Fourier transform (GFT) of a graph signal $\xb^{*}\in \mathbb{R} ^N$ is defined as its projection onto an orthogonal set of vectors $\{\ub_i\}_{i=1,...,N}$, i.e.

%%%%%%%%%%%%%%%%%%%%%%%%%%%%%%%%%%%%%%%%%%%%
\begin{equation}
\sbb^{*}=\Ub^H \xb^{*};\;
\xb^{*}=\Ub \sbb^{*}.
\end{equation}
%%%%%%%%%%%%%%%%%%%%%%%%%%%%%%%%%%%%%%%%%%%%
\noindent A bandlimited graph signal is defined as a graph signal whose GFT contains non-zero components only up to a specified bandwidth $\mathrm{BW}$. Formally, the signal is said to be bandlimited if the maximum index of its non-zero GFT coefficients does not exceed $\mathrm{BW}$. A bandlimited graph signal is inherently smooth, and this type of signal is assumed throughout this work. We consider the presence of adversarial nodes in the network. Specifically, a random number $K$ of nodes are assumed to behave adversarially by reporting erroneous values instead of their true signal components. Let $\xb^*$ denote the true graph signal. The observed signal $\xb$ is then modeled as
%%%%%%%%%%%%%%%%%%%%%%%%%%%%%%%%%%%%%%%%%%%%
\begin{align}
	\xb = (\Ib - \Mb_a)(\xb^* + \vb) + \Mb_a \xb_a,
\end{align}
%%%%%%%%%%%%%%%%%%%%%%%%%%%%%%%%%%%%%%%%%%%%%
 where $\xb_a$ is the adversarial signal vector, $\Mb_a$ is a diagonal Bernoulli masking matrix whose diagonal elements are drawn independent and identically distributed (i.i.d.) from a Bernoulli distribution with parameter $p_a$, indicating whether a node is attacked ($1$) or not ($0$), $\vb$ is the additive measurement noise vector, where each element $v_i$ is an i.i.d. zero-mean Gaussian random variable with variance $\sigma_\nu^2$, and $\xb_a$ contains adversarial values modeled as i.i.d. zero-mean Gaussian random variables with variance $\sigma_a^2$. The goal is to recover the true graph signal $\xb^*$ in a blind setting, where the identities of the adversarial nodes are unknown. A secondary objective is the detection of adversarial nodes.

\section{Proposed Method for Adversary Detection and Recovery}
\label{sec: prop}
\noindent In this section, a method to detect adversarial nodes and subsequently reconstruct the graph signal is proposed using a smoothness-based criterion.

\subsection{Adversary Detection in GSR}
\noindent In here, a closed-form, low-complexity adversary detection method is introduced, based on the smoothness variation of the graph signal. The central idea is that removing or correcting a signal component at an adversarial node should significantly change the smoothness of the signal. Let us define the smoothness of a graph signal $\xb$ as $	\mathrm{S}(\xb) = \xb^\top \Lb \xb$, where $\Lb$ denotes the graph Laplacian matrix and the \emph{minus-$k$ signal} as $\xb^{\langle -k \rangle} \triangleq \xb + (x_k - x_k^0)\eb_k$, where $\eb_k$ is the one-hot vector with a $1$ at index $k$, $x_k^0$ is the observed signal value at node $k$, and $x_k$ is a substitute (nominal adversary) value for $x_k$.

\noindent The proposed adversary detection criterion at node $k$ is given by:
%%%%%%%%%%%%%%%%%%%%%%%%%%%%%%%%%%%%%%%%%%%%
\begin{align}
	\mathrm{H}_{\text{adv-det-}k} \triangleq \mathrm{S}(\xb) - \mathbb{E}_{x_k}\left\{\mathrm{S}(\xb^{\langle -k \rangle})+ \lambda_k (x_k - x_k^0)^2\right\} ,
\end{align}
%%%%%%%%%%%%%%%%%%%%%%%%%%%%%%%%%%%%%%%%%%%%
\noindent where $\mathbb{E}_{x_k}\{\cdot\}$ is the expectation over the random variable $x_k$, and $\lambda_k > 0$ is a regularization parameter that encourages deviation from the expected value.

\noindent To derive a more tractable form, we define:
%%%%%%%%%%%%%%%%%%%%%%%%%%%%%%%%%%%%%%%%%%%%
\begin{align}
	\Delta_k &\triangleq \xb^\top \Lb \xb - {\xb^{\langle -k \rangle}}^\top \Lb \xb^{\langle -k \rangle} + \lambda_k (x_k - x_k^0)^2 \nonumber\\
	&= \xb^\top \Lb \xb - \left[\xb + \eb_k (x_k - x_k^0)\right]^\top \Lb \left[\xb + \eb_k (x_k - x_k^0)\right] \nonumber\\
	&\hspace{10mm} + \lambda_k (x_k - x_k^0)^2 \nonumber\\
	&= -2 \xb^\top \Lb \eb_k (x_k - x_k^0) - (\eb_k^\top \Lb \eb_k - \lambda_k)(x_k - x_k^0)^2.
\end{align}
%%%%%%%%%%%%%%%%%%%%%%%%%%%%%%%%%%%%%%%%%%%%
\noindent Let $\mathrm{H}_k = \mathbb{E}_{x_k}[\Delta_k]$. Assume $x_k \sim \mathcal{N}(0, \sigma^2)$, and define $\eta_k \triangleq L_{kk} - \lambda_k$, where $L_{kk}$ is the $k$-th diagonal element of $\Lb$, and $\Lb_k \triangleq \Lb \eb_k$ (i.e., the $k$-th column of the Laplacian). We then have:

%%%%%%%%%%%%%%%%%%%%%%%%%%%%%%%%%%%%%%%%%%%%
\begin{align}
	\mathrm{H}_k &= -\mathbb{E}\left\{2 \xb^\top \Lb_k (x_k - x_k^0) - \eta_k (x_k - x_k^0)^2 \right\} \nonumber\\
	&= 2 \xb^\top \Lb_k x_k^0 + \eta_k \left( \mathbb{E}[x_k^2] + (x_k^0)^2 \right) \nonumber\\
	&= 2 a_k x_k^0 + \eta_k \left(\sigma^2 + (x_k^0)^2 \right),
\end{align}
%%%%%%%%%%%%%%%%%%%%%%%%%%%%%%%%%%%%%%%%%%%%
\noindent where $a_k \triangleq \xb^\top \Lb_k$. The final detection criterion can then be written as:
%%%%%%%%%%%%%%%%%%%%%%%%%%%%%%%%%%%%%%%%%%%%
\begin{align}
	\mathrm{T}_f = \frac{2 a_k x_k^0}{\eta_k} - (x_k^0)^2 > \mathrm{Th} = \sigma^2.
\end{align}
%%%%%%%%%%%%%%%%%%%%%%%%%%%%%%%%%%%%%%%%%%%%
\noindent This expression provides a computationally efficient, closed-form test statistic for detecting adversarial nodes. The threshold $\mathrm{Th}$ can be set based on the noise variance $\sigma^2$. Simulation results (shown in Section~\ref{sec:sim}) demonstrate the effectiveness of this detector. In addition, an analytical expression for the optimal value of $\eta_k$ will be derived.

\subsection{GSR using the detected adversaries}
\noindent Consider $\xb \in \mathbb{R}^N$ as a received vector and $\Mb \in \mathbb{R}^{N \times N}$ is a binary diagonal detected attacks matrix (that is, $\Mb = \operatorname{diag}(m_1, \dots, m_N)$, where $m_i \in \{0,1\}$) where $m_i=1$ means that the $i$'th node is adversary and $m_i=0$ shows that the $i$'th node is honest and is only contaminated by noise.

\noindent To define a cost function based on normalized graph signal smoothness, we need to formulate the following optimization problem:
%%%%%%%%%%%%%%%%%%%%%%%%%%%%%%%%%%%%%%%%%%%%
\begin{equation}
\min_{\ub \in \mathbb{R}^N} \quad \frac{\left((\Ib - \Mb)\xb + \Mb\ub\right)^\top \Lb \left((\Ib - \Mb)\xb + \Mb\ub\right)}{\xb^\top (\Ib -\Mb)\xb + \ub^\top \Mb\ub}
\end{equation}
%%%%%%%%%%%%%%%%%%%%%%%%%%%%%%%%%%%%%%%%%%%%
\noindent such that $(\Ib -\Mb)\xb = (\Ib -\Mb)\ub$.

\noindent Reformulation:

Let:
\begin{itemize}
	\item $\mathcal{I}_x = \{ i : m_i = 0 \}$,
	\item $\mathcal{I}_u = \{ i : m_i = 1 \}$,
	\item $\zb := \ub_{\mathcal{I}_u} \in \mathbb{R}^{|\mathcal{I}_u|}$,
	\item $\xb_0  := \ub_{\mathcal{I}_x} \in \mathbb{R}^{|\mathcal{I}_x|}$.
	\end{itemize}
\noindent Reorder $\Lb$ accordingly:
%%%%%%%%%%%%%%%%%%%%%%%%%%%%%%%%%%%%%%%%%%%%
\begin{equation}
\hat{\Lb} \triangleq
\begin{bmatrix}
	\Lb_{xx} & \Lb_{xu} \\
	\Lb_{ux} & \Lb_{uu}
\end{bmatrix}
\end{equation}
%%%%%%%%%%%%%%%%%%%%%%%%%%%%%%%%%%%%%%%%%%%%
\noindent Then the numerator becomes:
%%%%%%%%%%%%%%%%%%%%%%%%%%%%%%%%%%%%%%%%%%%%
\begin{align}
f(\zb) &=
\begin{bmatrix}
	\xb_0 \\ \zb
\end{bmatrix}^\top
\begin{bmatrix}
	\Lb_{xx} & \Lb_{xu} \\
	\Lb_{ux} & \Lb_{uu}
\end{bmatrix}
\begin{bmatrix}
	\xb_0 \\ \zb
\end{bmatrix}\notag\\
&= \xb_0^\top \Lb_{xx} \xb_0 + 2 \zb^\top \Lb_{ux} \xb_0 + \zb^\top \Lb_{uu} \zb
\end{align}
%%%%%%%%%%%%%%%%%%%%%%%%%%%%%%%%%%%%%%%%%%%%
\noindent The denominator becomes $g(\zb) = \xb_0^\top \xb_0 + \zb^\top \zb$. So the fractional objective is:
%%%%%%%%%%%%%%%%%%%%%%%%%%%%%%%%%%%%%%%%%%%%
\begin{equation}
\label{eq: opt}
\min_{\zb \in \mathbb{R}^{|\mathcal{I}_x|}} \quad R(\zb) \triangleq \frac{f(\zb)}{g(\zb)} =\frac{\zb^\top \Bb \zb + 2 \bb^\top\zb + \alpha}{\zb^\top \zb + \beta},
\end{equation}
%%%%%%%%%%%%%%%%%%%%%%%%%%%%%%%%%%%%%%%%%%%%
\noindent where it is a generalized Rayleigh quotient, with
\begin{itemize}
	\item $\Bb = \Lb_{uu}$,
	\item $\bb = \Lb_{ux} \xb_0$,
	\item $\alpha = \xb_0^\top \Lb_{xx} \xb_0\geq0$,
	\item $\beta = \xb_0^\top \xb_0\geq0$.
\end{itemize}
% $\Bb = \Lb_{uu}$, $\bb = \Lb_{ux} \xb_0$, $\alpha = \xb_0^\top \Lb_{xx} \xb_0\geq0$, and $\beta = \xb_0^\top \xb_0\geq0$.
 The fractional optimization problem in (\ref{eq: opt}) can be solved via a Dinkelbach's Algorithm which is outlined as:
\\\\
\noindent \textbf{Dinkelbach's Algorithm:}

\noindent Initialize $\gamma^{(0)}$. Iterate until convergence:

1. Solve:
\begin{align}
	\zb^{(\ell)}  &= \arg\min_{\zb \in \mathbb{R}^{|\mathcal{I}_x|}} \quad Q(\zb);\\
	Q_{\gamma^{(\ell)} }(\zb)&\triangleq{\zb^\top \Bb \zb + 2 \bb^\top\zb + \alpha} - \gamma^{(\ell)} ({\zb^\top \zb + \beta})
\end{align}

2. Update:

\begin{equation}
\gamma^{(\ell+1)} = \frac{{(\zb^{(\ell)})^\top \Bb \zb^{(\ell)} + 2 \bb^\top \zb^{(\ell)} + \alpha}}{{(\zb^{(\ell)})^\top \zb^{(\ell)} + \beta}}.
\end{equation}

3. Check convergence $\left| f(\zb^{(\ell)}) - \gamma^{(\ell)} g(\zb^{(\ell)}) \right| < \epsilon$.

\noindent Let $\zb^*$ be the minimizer at convergence. Construct the final solution $\hat{\xb} \in \mathbb{R}^n$ as $	\hat{\xb} = \ub^* = \Mb.\zb + (\Ib-\Mb).\xb$. For a proply chosen regularization parameter $\gamma$, the gradient of the objective function is given by
%%%%%%%%%%%%%%%%%%%%%%%%%%%%%%%%%%%%%%%%%%%%%
\begin{align}
	\nabla_\zb Q_{\gamma}(\zb) = \nabla_\zb \left[ f(\zb) - \gamma g(\zb) \right] = 2\Bb \zb + 2\bb - 2\gamma \zb,
\end{align}
%%%%%%%%%%%%%%%%%%%%%%%%%%%%%%%%%%%%%%%%%%%%
where $\Bb$ is a symmetric matrix, $\bb$ is a vector, and $\zb$ is the optimization variable. This gradient expression can be used to perform iterative updates to minimize $Q_\gamma(\zb)$ using gradient descent. Justifications for Using Dinkelbach's Algorithm are as follows:

1. Convexity of $ f(\zb) $
\newline Since $ \Lb \succeq 0 $, the quadratic form $ \zb^\top \Lb \zb $ is convex. The affine term $ 2\bb^\top \zb $ preserves convexity. Hence, $ f(\zb) $ is convex.

2. Convexity and positivity of $ g(\zb) $
\newline$g(\zb) = \zb^\top \zb + \beta
$ is strictly convex (positive definite quadratic plus constant) and strictly positive for all $ {\zb \in \mathbb{R}^{|\mathcal{I}_x|}} $ due to $ \beta > 0 $.

3. Pseudo-Convexity of $ R(\zb) $
\newline If $ f $ is convex, $ g $ is strictly positive and convex, and both are differentiable, then $ R(\zb) = f(\zb)/g(\zb) $ is pseudo-convex~\cite{schaible1976fractional,ben2001lectures}. Thus, any stationary point is a global minimizer, and Dinkelbach's algorithm converges globally \cite{dinkelbach1967nonlinear}.

-Iterative Solver: MINRES:
The minimizer of the quadratic form satisfies:
%%%%%%%%%%%%%%%%%%%%%%%%%%%%%%%%%%%%%%%%%%%%
\begin{align}
&\nabla Q_{\gamma^{(\ell)} }(\zb) = 2(\Lb - \gamma^{(\ell)}  \Ib)\zb + 2\bb = 0 \notag\\
& \Rightarrow \quad (\Lb - \gamma^{(\ell)}  \Ib) \zb = -\bb.
\end{align}
%%%%%%%%%%%%%%%%%%%%%%%%%%%%%%%%%%%%%%%%%%%%
\noindent Since $ \Lb - \gamma^{(\ell)} \Ib $ is symmetric and may be indefinite, we use the Minimum Residual method (MINRES) algorithm ~\cite{paige1975solution}, which:
\begin{itemize}
	\item Solves symmetric (possibly indefinite as in our case) linear systems,
	\item Avoids direct matrix inversion,
	\item Is suitable for large, sparse systems like those from Laplacian matrices.
\end{itemize}
Hence, Dinkelbach's algorithm combined with MINRES is both theoretically sound and practically efficient for solving the given generalized Rayleigh quotient minimization.

\section{Analysis and the Detector Algorithms}
\label{sec: ana}
\noindent In this section, we calculate the Missed Detection (MD) and False Detection (FD) probabilities. For each node $k$, they are defined as $\mathrm{P}_\text{err, k}\triangleq \mathrm{P}_\text{MD, k} + \mathrm{P}_\text{FD, k}$, $\mathrm{P}_\text{MD, k}\triangleq \mathrm{Pr}\{\mathrm{T}_{\mathrm{f}}<\mathrm{Th}|\text{Adversary}\}$, and $\mathrm{P}_\text{FD, k}\triangleq \mathrm{Pr}\{\mathrm{T}_{\mathrm{f}}>\mathrm{Th}|\text{No Adversary}\}$. To calculate the detection probability, the final detection statistics can be rewritten as
%%%%%%%%%%%%%%%%%%%%%%%%%%%%%%%%%%%%%%%%%%%%
\begin{align}
\mathrm{T}_{\mathrm{f}}&=2\frac{\Big(\sum_ix^0_iL_{ki}\Big)x^0_k}{\eta_k}-{x_k^0}^2=c{x_k^0}^2+dx^0_k,
\end{align}
%%%%%%%%%%%%%%%%%%%%%%%%%%%%%%%%%%%%%%%%%%%%
\noindent where $c\triangleq 2\frac{L_{kk}}{\eta_k}-1$, and $d\triangleq 2\frac{\sum_{i\neq k}x^0_iL_{ki}}{\eta_k}$. By making the statistics as a complete square term, we have
%%%%%%%%%%%%%%%%%%%%%%%%%%%%%%%%%%%%%%%%%%%%
\begin{align}
\mathrm{T}_{\mathrm{f}}=c(x^0_k+e)^2+f,
\end{align}
%%%%%%%%%%%%%%%%%%%%%%%%%%%%%%%%%%%%%%%%%%%%
\noindent where $e=\frac{d}{2c}$ and $f=-ce^2$. For tractable analysis, we assume the Gaussianity and zero mean of $x^0_k$ for the adversary signal which is $x^0_k\sim N(0,\sigma^2)$. So, the term $x^0_k+e\sim N(e,\sigma^2)$. Hence, we have
%%%%%%%%%%%%%%%%%%%%%%%%%%%%%%%%%%%%%%%%%%%%
\begin{align}
	\mathrm{P}_\text{MD, k}&= \mathrm{Pr}\{  |x^0_k+e|<\theta |\text{Adversary}\}\notag\\
	&= 1 - Q(\frac{\theta+\mu_m}{\sigma_m} )- Q(\frac{\theta-\mu_m}{\sigma_m})\\
	\mathrm{P}_\text{FD, k}&=\mathrm{Pr}\{|x^0_k+e|>\theta|\text{No Adversary}\}\notag\\
	&= Q(\frac{\theta+\mu_f}{\sigma_f} )+ Q(\frac{\theta-\mu_f}{\sigma_f}) .
\end{align}
%%%%%%%%%%%%%%%%%%%%%%%%%%%%%%%%%%%%%%%%%%%%
\noindent where
$\theta = \sqrt{\frac{\mathrm{Th-f}}{c}}$, $\mu_m=e$, $\mu_f = \mu_m + x^*_k$, $\sigma_m = \sigma_a$, $\sigma_f = \sigma_\nu$.

\noindent We should select $\eta_k<2L_{kk}$ to ensure $c>0$.

\noindent The optimization problem is formulated as
%%%%%%%%%%%%%%%%%%%%%%%%%%%%%%%%%%%%%%%%%%%%
\begin{equation}
	\min Q(\frac{\theta+\mu_f}{\sigma_\nu} )+ Q(\frac{\theta-\mu_f}{\sigma_\nu})- Q(\frac{\theta+\mu_m}{\sigma_a} )- Q(\frac{\theta-\mu_m}{\sigma_a})
	\label{eq:eta_opt_problem}
\end{equation}
%%%%%%%%%%%%%%%%%%%%%%%%%%%%%%%%%%%%%%%%%%%%
\noindent In order to find the optimal threshold level $\mathrm{Th}$, one can numerically minimize the detection error probability $\mathrm{P}_{\text{err},k}$.
Alternatively, a more analytical approach involves identifying the zero-crossing of the derivative of $\mathrm{P}_{\text{err},k}$ with respect to $\mathrm{Th}$, i.e., solving:
\begin{align}
	&\frac{\partial \mathrm{P}_\text{err, k}}{\partial \mathrm{Th}} = \frac{\partial \mathrm{P}_\text{MD, k}}{\partial \mathrm{Th}} +\frac{\partial \mathrm{P}_\text{FD, k}}{\partial \mathrm{Th}} \notag\\
	&=   \frac{\partial\theta }{\partial \mathrm{Th}}.\Big(\frac{\Phi(\frac{\theta+\mu_m}{\sigma_a}) + \Phi(\frac{\theta-\mu_m}{\sigma_a})}{\sigma_a}-\frac{\Phi(\frac{\theta+\mu_f}{\sigma_\nu} )+ \Phi(\frac{\theta-\mu_f}{\sigma_\nu})}{\sigma_\nu}\Big),
\end{align}

\noindent where $\Phi(.)$ is the probability distribution function (PDF) of the standard normal distribution and we have $\frac{\partial\theta }{\partial \mathrm{Th}} =  \frac{1}{2c\theta}$. To solve the optimization problem in \eqref{eq:eta_opt_problem}, we aim to find the optimal threshold level $\mathrm{Th}$ that minimizes the overall error probability $\mathrm{P}_\text{err, k}$. Since the objective function is smooth, continuous, and unimodal in $\mathrm{Th}$ within a practical range, derivative-free optimization techniques are suitable and efficient. Specifically, we employ the \textit{golden section search} and \textit{parabolic interpolation} methods, which are classical one-dimensional optimization algorithms. The golden section search is robust and guarantees convergence by progressively narrowing the search interval, even in the absence or complexity of derivative information. On the other hand, the parabolic method leverages local curvature information to accelerate convergence by fitting a parabola through sampled points. Combining both methods provides a balanced approach: golden search ensures global convergence in the early phase, while parabolic interpolation refines the solution in the final stage with improved precision and faster convergence. This hybrid approach is computationally efficient and well-suited to our problem structure, where analytic gradients are either unavailable or computationally expensive to evaluate. The same approach can be followed to determine the optimal regularization parameter $\eta$. The derivative of the detection error probability with respect to $\eta$ is given by:
%%%%%%%%%%%%%%%%%%%%%%%%%%%%%%%%%%%%%%%%%%%%
\begin{align}
	 &\frac{\partial \mathrm{P}_\text{err, k}}{\partial \eta} = \frac{\partial \mathrm{P}_\text{MD, k}}{\partial \eta} +\frac{\partial \mathrm{P}_\text{FD, k}}{\partial \eta} \\
	 &=  \frac{ \frac{\partial\theta }{\partial \eta}+\frac{\mu_a }{\partial \eta}}{\sigma_a}.\Phi(\frac{\theta+\mu_m}{\sigma_a}) +\frac{ \frac{\partial\theta }{\partial \eta}-\frac{\mu_m }{\partial \eta}}{\sigma_a}. \Phi(\frac{\theta-\mu_m}{\sigma_a})\notag\\
	  &  -\frac{ \frac{\partial\theta }{\partial \eta}+\frac{\mu_f }{\partial \eta}}{\sigma_\nu}.\Phi(\frac{\theta+\mu_f}{\sigma_\nu} )-  \frac{ \frac{\partial\theta }{\partial \eta}-\frac{\mu_f }{\partial \eta}}{\sigma_\nu}.\Phi(\frac{\theta-\mu_f}{\sigma_\nu})\\
&=     \frac{\partial\mu_m }{\partial \eta}.\Big(\frac{\Phi(\frac{\theta+\mu_m}{\sigma_a}) - \Phi(\frac{\theta-\mu_m}{\sigma_a})}{\sigma_a}-\frac{\Phi(\frac{\theta+\mu_f}{\sigma_\nu} )- \Phi(\frac{\theta-\mu_f}{\sigma_\nu})}{\sigma_\nu}\Big)\notag\\
&+\frac{\partial\theta }{\partial\eta}.\Big(\frac{\Phi(\frac{\theta+\mu_m}{\sigma_a}) + \Phi(\frac{\theta-\mu_m}{\sigma_a})}{\sigma_a}-\frac{\Phi(\frac{\theta+\mu_f}{\sigma_\nu} )+ \Phi(\frac{\theta-\mu_f}{\sigma_\nu})}{\sigma_\nu}\Big) ,
\label{eq:opt_et_derv}
\end{align}
%%%%%%%%%%%%%%%%%%%%%%%%%%%%%%%%%%%%%%%%%%%%
\noindent where, for simplicity, we denote $\eta = \eta_k$. The partial derivatives involved are given by:
%%%%%%%%%%%%%%%%%%%%%%%%%%%%%%%%%%%%%%%%%%%%
\begin{align}
	 \frac{\partial\theta }{\partial \eta} = \frac{2c(c+1)\mathrm{Th}-d^2}{\eta c^3\theta};\,\frac{\partial\mu_m }{\partial \eta}=\frac{\partial\mu_f }{\partial \eta} =\frac{\eta\mu_m}{c}.
\end{align}
%%%%%%%%%%%%%%%%%%%%%%%%%%%%%%%%%%%%%%%%%%%%
\noindent Note that, due to the inaccessibility of the true value $x^*_k$ and the smoothness assumption of the graph signal, the component $x^*_k$ used in computing $\mu_f$ can be approximated as $\sum_ix^0_i/ N$. Solving \eqref{eq:opt_et_derv} yields the optimal value of $\eta_k$ for a given threshold $\mathrm{Th}$.

%%%%%%%%%%%%%%%%%%%%%%%%%%%%%%%%%%%%%%%%%%%%
\begin{figure}[!t]
\begin{center}
\includegraphics[width=9cm, height=6cm]{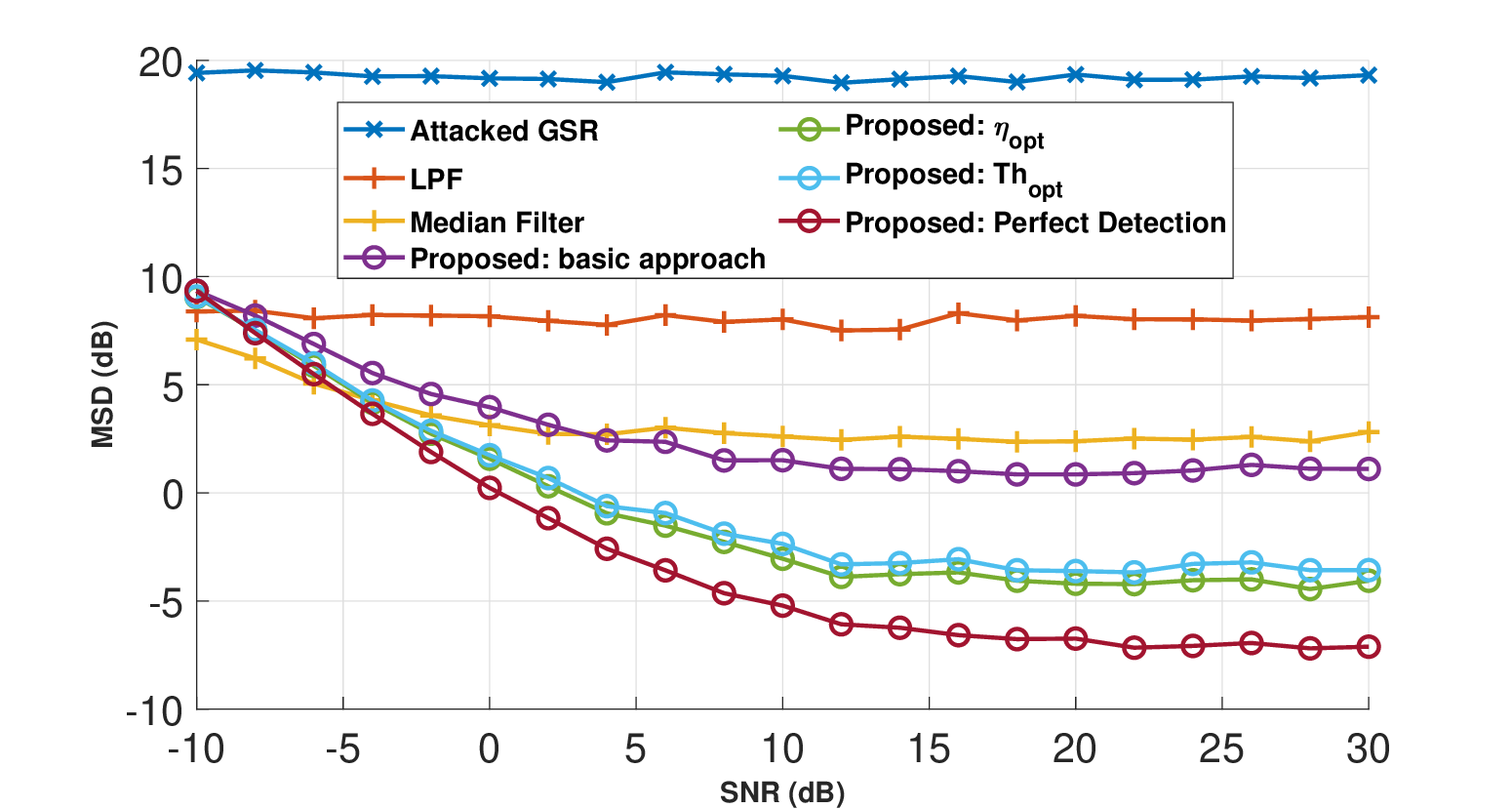}
\end{center}
\caption{Performance comparison using adversary detection.}
\label{fig:comp_th}
\end{figure}
%%%%%%%%%%%%%%%%%%%%%%%%%%%%%%%%%%%%%%%%%%%%
%%%%%%%%%%%%%%%%%%%%%%%%%%%%%%%%%%%%%%%%%%%%
\begin{figure}[!t]
\begin{center}
\includegraphics[width=9cm, height=6cm]{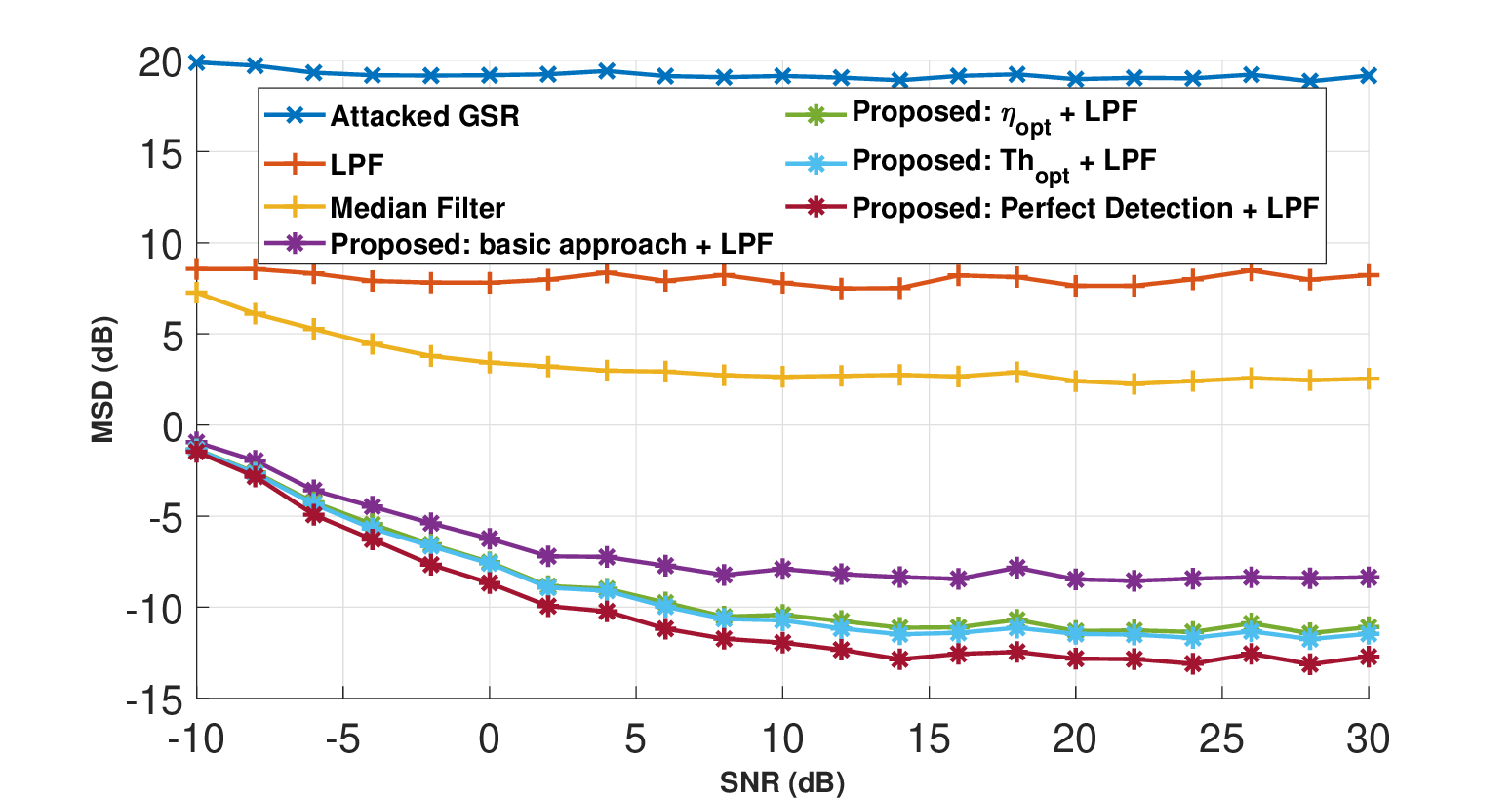}
\end{center}
\caption{Performance comparison with exploiting smoothness and applying an LPF.}
\label{fig:comp_lpf}
\end{figure}
%%%%%%%%%%%%%%%%%%%%%%%%%%%%%%%%%%%%%%%%%%%%
\section{Simulation Results}
\label{sec:sim}
\noindent In this section, simulation results are presented under the following setup. An Erdos-Renyi graph \cite{GSPbook_ortega} is generated with $N=20$ nodes and a connection probability of $p_{\text{link}} = 0.3$, which ensures connectivity. Edge weights are drawn uniformly at random from the interval $[0.5, 1]$. A normalized graph signal is randomly generated and designed to be smooth in the graph spectral domain by retaining only $\mathrm{BW}=2$ low-frequency components in the GFT. The observed signal is then contaminated with i.i.d. Gaussian noise, with the noise level adjusted to match a specified signal-to-noise ratio (SNR). Each node is independently attacked with probability $p_a = 0.2$, where the adversarial noise follows a Gaussian distribution with $\sigma_a = 5$. Results are averaged over 1000 Monte Carlo realizations. Performance is measured in terms of the Mean Square Deviation (MSD), defined as $	\mathrm{MSD} = \frac{\|\hat{\xb} - \xb^*\|_2}{\|\xb^*\|_2}$, where $\hat{\xb}$ denotes the estimated signal and $\xb^*$ is the ground truth.
%%%%%%%%%%%%%%%%%%%%%%%%%%%%%%%%%%%%%%%%%%%%
\begin{figure}[!t]
\begin{center}
\includegraphics[width=9cm, height=6cm]{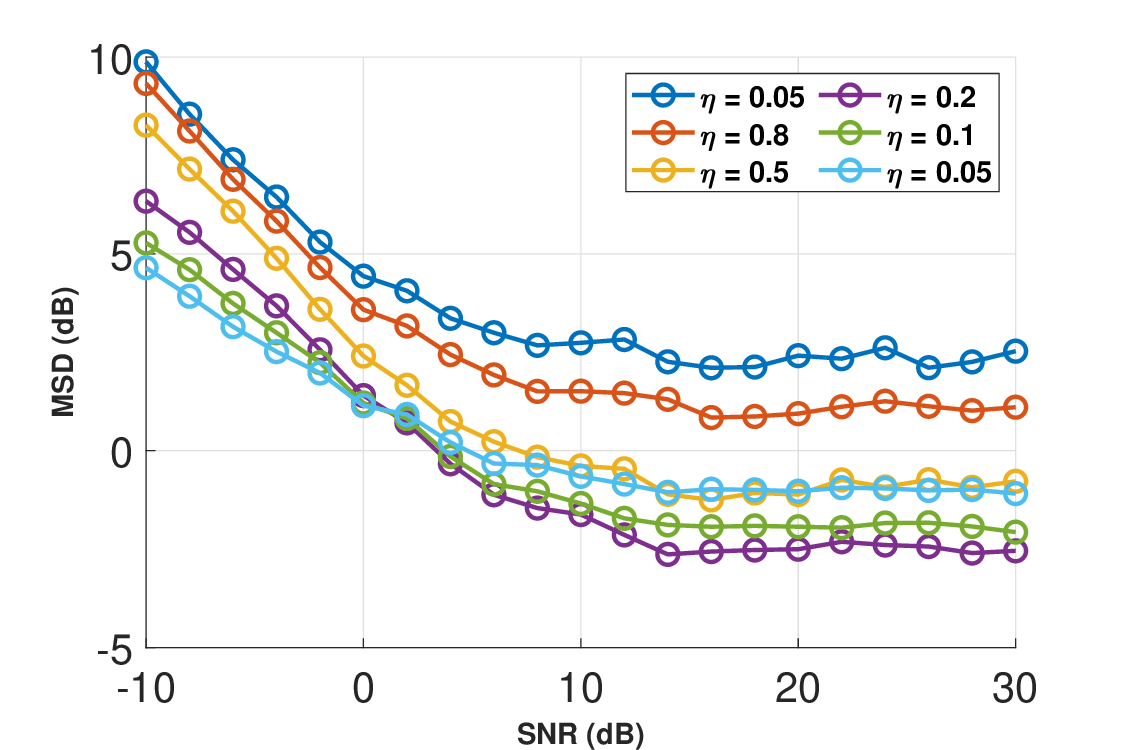}
\end{center}
\caption{Performance comparison using different $\eta$ values.}
\label{fig:comp_eta}
\end{figure}
%%%%%%%%%%%%%%%%%%%%%%%%%%%%%%%%%%%%%%%%%%%
\noindent Figure~\ref{fig:comp_th} illustrates the performance in the first scenario. The proposed approach is evaluated with the fixed parameters $\eta = 0.8$ and $\mathrm{Th} = \sigma_a^2$, referred to as the ``basic approach". For comparison, results using the parameters $\eta_{\text{opt}}$ and $\mathrm{Th}_{\text{opt}}$, obtained by solving the optimization problem in \eqref{eq:eta_opt_problem}, are also included. Additionally, an idealized case with perfect adversary detection is considered as a performance benchmark. The results show that the proposed method significantly improves the robustness of GSR under attack, achieving an MSD improvement of approximately 24\,dB compared to the attacked scenario. Furthermore, it outperforms traditional methods such as Low-Pass Filtering (LPF) and Median Filtering by margins of about 12\,dB and 6\,dB, respectively. Considering the smoothness of the signal, an LPF can be applied to the reconstructed signal as a post-processing step. As illustrated in Fig.~\ref{fig:comp_lpf}, the performance of the attacked Graph Signal Recovery (GSR) improves by approximately 30\,dB after applying the proposed method. Furthermore, the adversary detection and recovery strategy can improve the performance of the LPF by an additional 20\ dB.
\noindent To investigate the impact of the parameter $\eta$, Fig.\ref{fig:comp_eta} shows the results for different values of $\eta$. It shows that the best value of $\eta$ in terms of reducing the final MSD is equal to $\eta=0.2$.

\section{Conclusion}
\label{sec: con}
\noindent In this paper, the problem of GSR in the presence of malicious nodes that inject FDI error is investigated. In addition to GSR, a low complexity adversary detector is presented based on the change in the smoothness of the observed graph signal and average replaced graph signal. After adversary detection, the GSR is performed via a smoothness maximization approach which is solved using an efficient iterative fractional optimization method. For the detector, the detection and false alarm probabilities are calculated in a closed form. Simulation results show that the proposed method achieves a performance gain of approximately 8\ dB in signal recovery compared to the graph median filtering.

\end{document}